\documentclass[3p,times,procedia]{elsarticle}

\usepackage{ecrc}

\volume{00}

\firstpage{1}

\journalname{Physics Procedia}

\runauth{A. Haungs}

\jid{phpro}

\jnltitlelogo{Physics Procedia}

\CopyrightLine{2014}{Published by Elsevier Ltd.}

\usepackage{amssymb}

\usepackage{lineno}

\usepackage[figuresright]{rotating}

\begin{document}

\begin{frontmatter}



\dochead{13th International Conference on Topics in Astroparticle and Underground Physics}

\title{Cosmic Rays from the Knee to the Ankle}


\author{Andreas Haungs}

\ead{haungs@kit.edu}

\address{Karlsruhe Institute of Technology (KIT), Institut f\"ur Kernphysik, 76021 Karlsruhe, Germany}

\begin{abstract}
Investigations of the energy spectrum as well as the mass composition of cosmic rays in the energy 
range of PeV to EeV are important for understanding both, the origin of the galactic and the extragalactic 
cosmic rays. 
Recently, three modern experimental installations (KASCADE-Grande, IceTop, Tunka-133),
dedicated to investigate this primary energy range, have published new results on the 
all-particle energy spectrum.
In this short review these results are presented and the similarities and differences discussed. 
In addition, the effects of using different hadronic interaction models for interpreting the 
measured air-shower data will be examined.
Finally, a brief discussion on the question if the present results are in agreement or in contradiction 
with astrophysical models for the transition from galactic to extragalactic origin of cosmic rays 
completes this paper.  
\end{abstract}


\end{frontmatter}


\section{Introduction}

Experimental cosmic ray research aims to determine the primary particle's arrival 
direction distribution, energy spectrum, and elemental composition. 
The measurements comprise important hints to gain knowledge in the origin, acceleration 
and propagation of these energetic particles of cosmic origin. 
At energies above $10^{15}\,$eV, the characteristics of these particles must be 
determined indirectly from the measured properties of extensive air showers (EAS) 
that cosmic rays induce in the Earth's atmosphere~\cite{Haungs}. 
 
The determination of the primary energy and elemental composition in the energy range 
from $10^{14}\,$eV up to above $10^{20}\,$eV is subject of air-shower experiments 
since more than six decades. 
It has been shown that the all-particle spectrum has a power-law like behavior 
($\propto\,E^{-\gamma}$, with $\gamma\,\approx\,2.7$)
with features, which are known as `knee' and `ankle' at 
$2$-$5\cdot10^{15}\,$eV and $2$-$8\cdot10^{18}\,$eV, respectively. 
Whereas at the knee the spectrum steepens to $\gamma\,\approx\,3.0$, the ankle is characterized by a flattening 
of the spectrum by roughly the same change of the spectral index; i.e.~back to $\gamma\,\approx\,2.7$.
Low-energy cosmic rays are of galactic origin and cosmic rays above the ankle are most probable of 
extragalactic origin~\cite{auger-eg}, i.e.~somewhere in the energy range from 
$10^{16}\,$eV to a few $10^{18}\,$eV 
the transition of cosmic rays of galactic to extragalactic origin is expected. 
\begin{figure}[ht]
\begin{center}
 \includegraphics[width=0.45\textwidth]{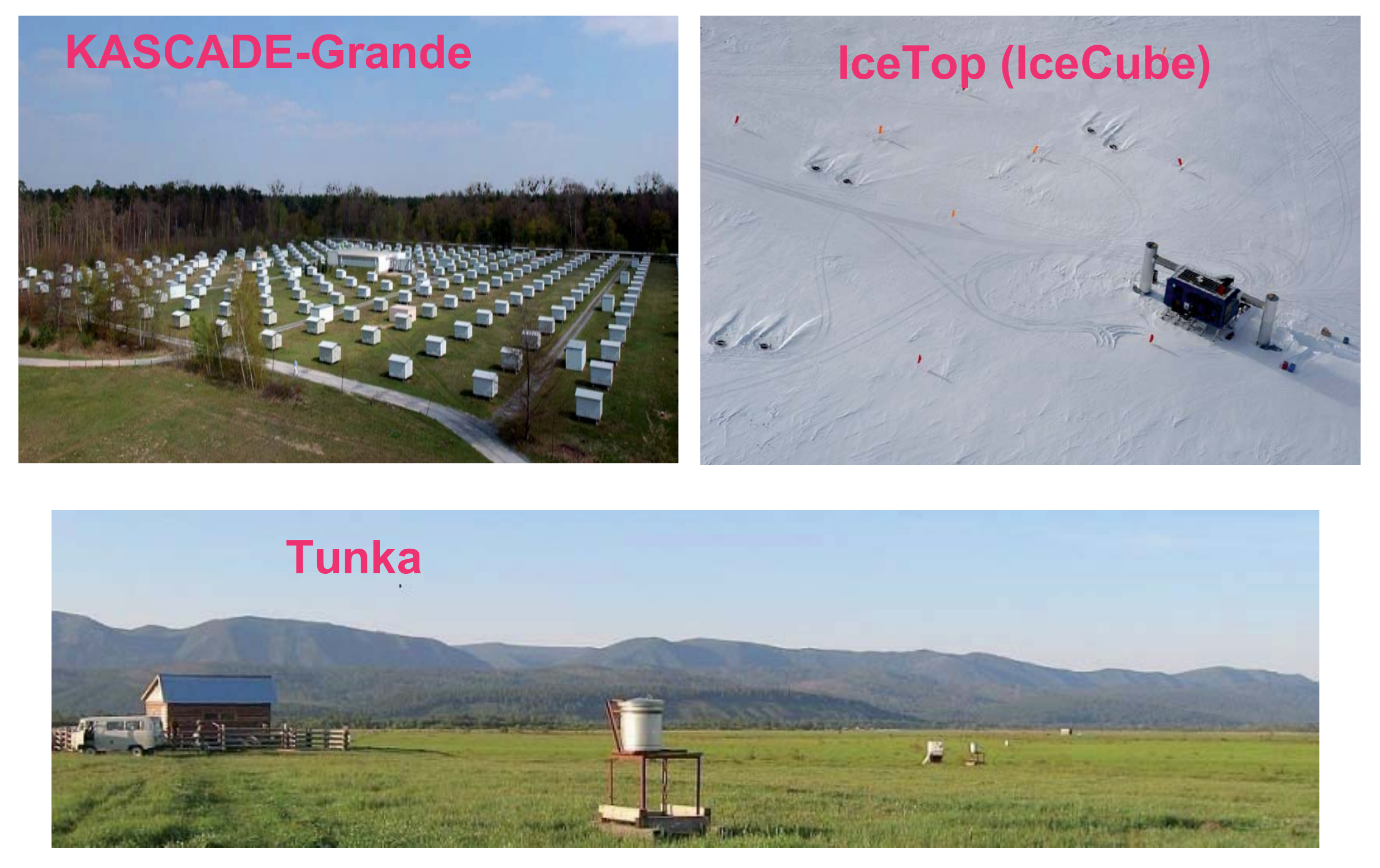} 
 \includegraphics[width=0.54\textwidth]{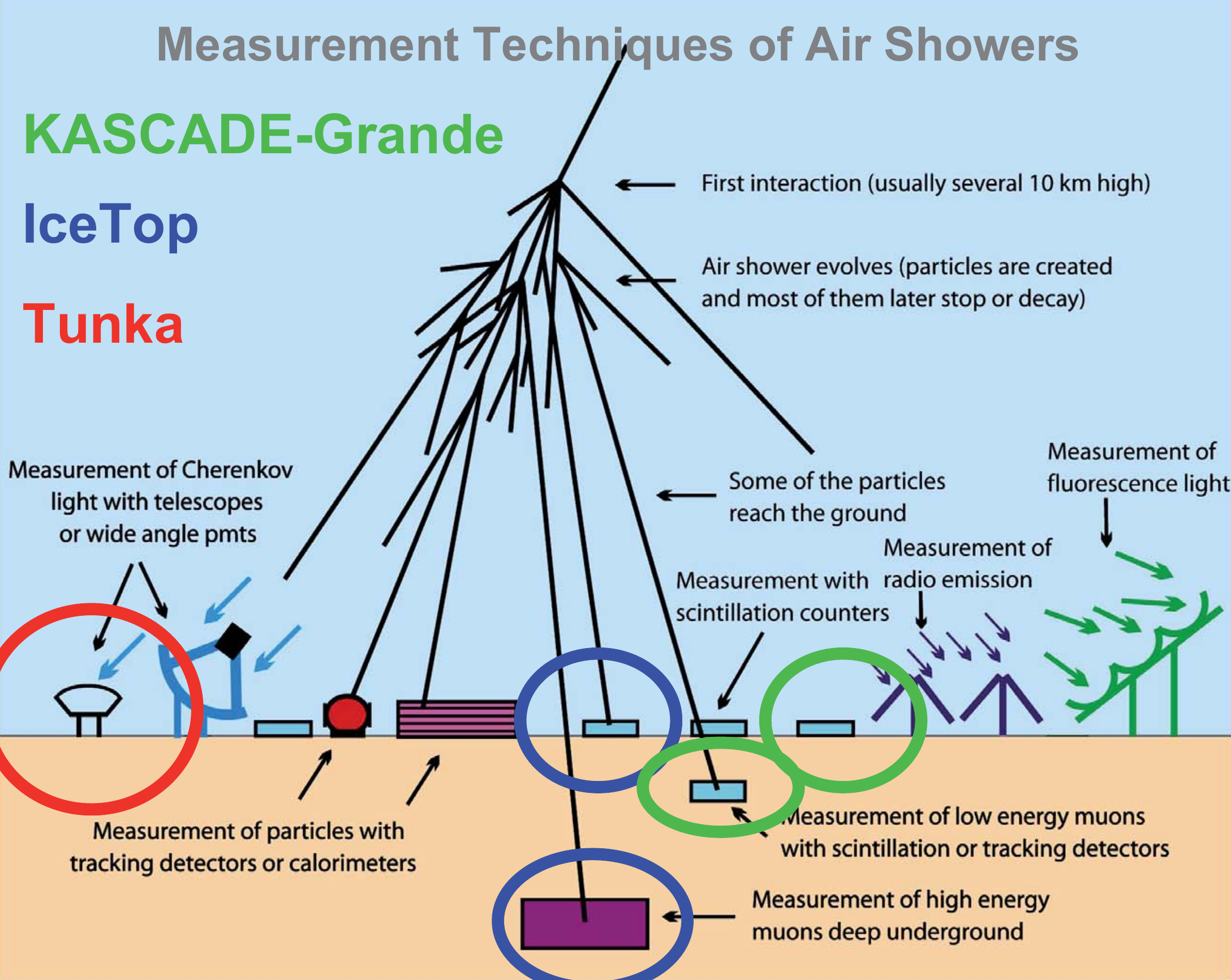} 
\end{center}
\caption{Photographies of the three air shower experiments KASCADE-Grande, IceTop, Tunka-133 (left).
Right panel: Sketch of the complementary methodical approaches for air shower measurements at 
the three experiments.}
\label{fig:meths}
\end{figure}

A decade ago, the KASCADE experiment~\cite{kascade} has shown that the knee feature is due to 
a distinct break in the intensity in the light component of cosmic rays 
(Z$<6$), only, where the difference in the energies of the knee features of protons and Helium nuclei 
is in agreement with the assumption of a rigidity (charge) dependent knee~\cite{kas-unf}.
It should be noted that protons are not the most abundant primary  
in this energy range, which is in agreement with extrapolations of proton and Helium spectra measured with 
the balloon experiment CREAM~\cite{cream}. 

Whereas the knee as well as the ankle experienced investigations by many experiments, the energy range 
$10^{16}$ - $10^{18}$\,eV stayed somehow unexplored until recently,  
though the study of primary energy spectrum and mass composition in this energy range 
is of crucial importance for understanding origin and propagation of cosmic rays. 
Meanwhile, three dedicated experiments presented new results for the all-particle energy spectrum, 
at least, in this energy range. 
In the present review we discuss these all-particle energy spectra of cosmic 
rays in the range from $10^{16}$ to $10^{18}\,$eV obtained by KASCADE-Grande~\cite{kg-NIM10}, 
Tunka-133~\cite{nim12}, and IceTop~\cite{IT_detP}; 
see Figure~\ref{fig:meths}, left panel.
Extensive air showers (EAS) are generated when high-energy cosmic particles enter the atmosphere. 
Forward-boosted secondary particles as well as emitted light during the development of the EAS in 
various frequency ranges form the detectable products; see Figure~\ref{fig:meths}, right panel.  
Depending on the experimental apparatus and the detection technique of ground-based 
air-shower experiments, different sets of EAS observables are available to estimate 
the energy of the primary cosmic rays~\cite{Haungs}. 

The three experiments under consideration are complementary in their detection method (Figure~\ref{fig:meths}). 
Whereas KASCADE-Grande measures 
the electromagnetic component with scintillators and IceTop by Ice-Cherenkov tanks, Tunka-133 
uses open Photomultipliers to measure the Cherenkov-light emitted by the electromagnetic component. 
KASCADE-Grande detects in addition low energy muons of the showers by shielded scintillators, and 
IceTop has the possibility to include the measurement of high-energy muons with IceCube, though not 
used yet for the determination of the all-particle energy spectrum. 

What follows is first a short description of the three experiments and their results before the differences 
and similarities of the spectra are discussed. In addition, a brief discussion on implications of the 
results for interpretation of the transition energy range is done.

\section{Experimental Results}

\subsection{KASCADE-Grande}
\begin{figure}[ht]
\begin{center}
 \includegraphics[width=0.52\textwidth]{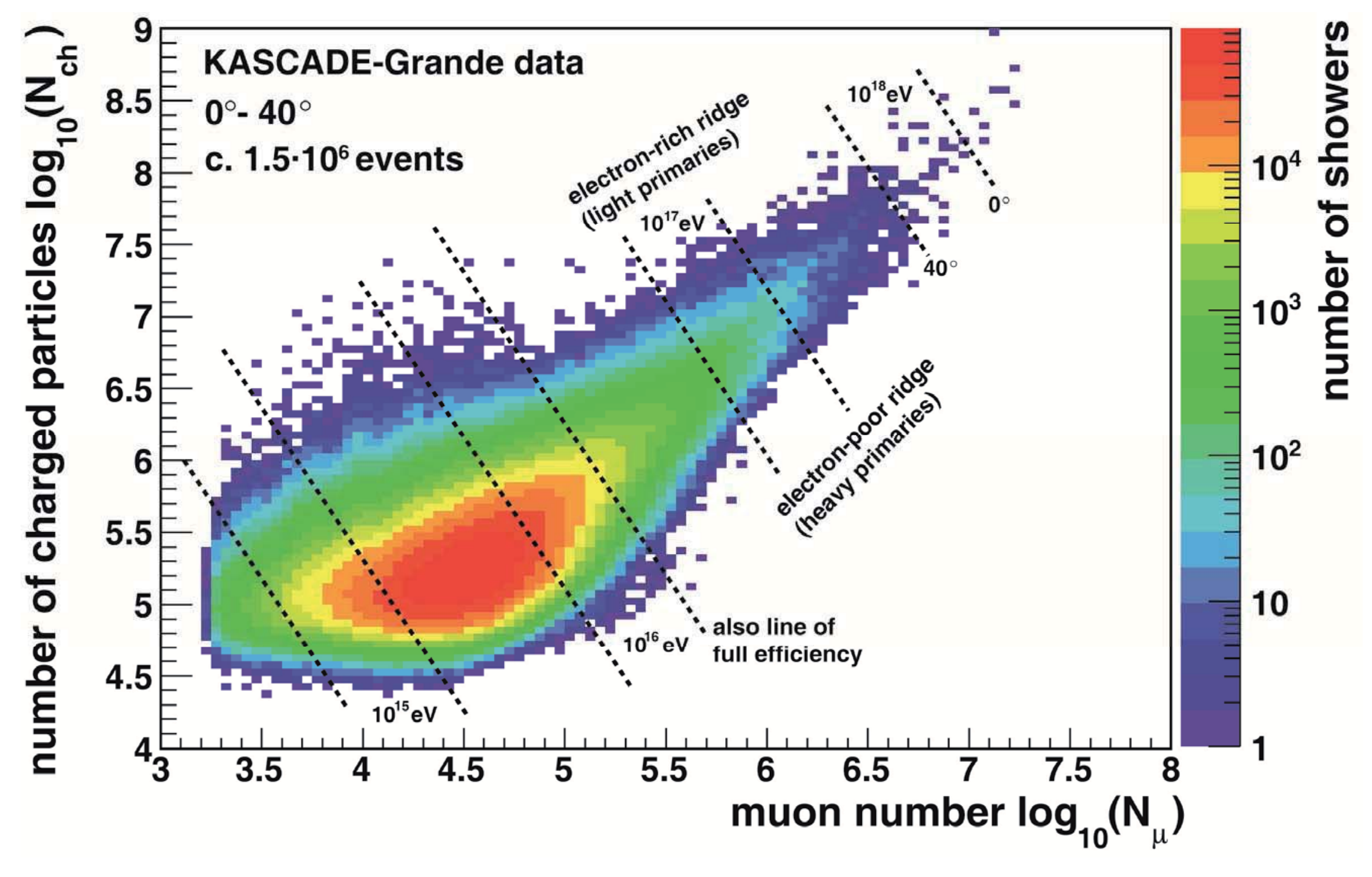} 
 \includegraphics[width=0.46\textwidth]{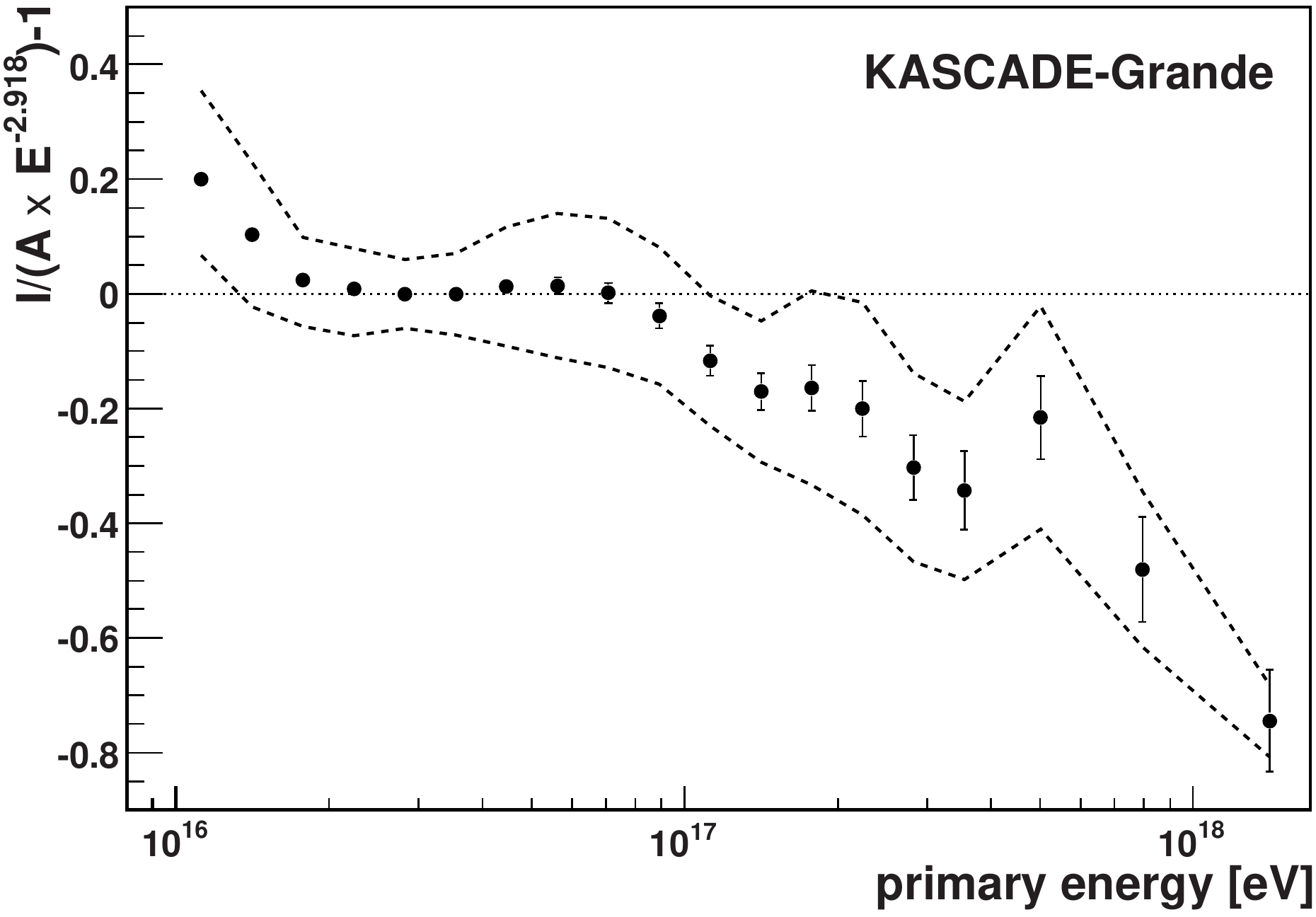} 
\end{center}
\caption{Left: Two-dimensional shower size spectrum measured by KASCADE (from~\cite{kgECRShaungs}).
Right: The all-particle energy spectrum obtained with KASCADE-Grande, where for a better visibility of the structures the
residual intensity I (including 
the band of systematic uncertainty) is displayed after multiplying the spectrum with a certain factor A (from~\cite{kgEspec}).}
\label{fig:kgspec}
\end{figure}
Main parts of KASCADE-Grande~\cite{kg-NIM10} are the Grande array spread over an 
area of $700 \times 700\,$m$^2$, 
the original KASCADE array covering $200 \times 200\,$m$^2$ with unshielded and shielded 
detectors, a large-size hadron calorimeter, and additional muon tracking devices. 
The estimation of energy and mass of the primary particles follows a combined 
investigation of the charged particle, the electron, and the muon components 
measured by the detector arrays of Grande and KASCADE. 
The multi-detector experiment KASCADE~\cite{kascade}
(located in Karlsruhe, Germany) was extended to KASCADE-Grande 
in 2003 by installing 37 additional stations consisting 
of 10$\,$m$^2$ scintillation detectors each.  
While the Grande detectors are sensitive to charged particles, 
the KASCADE array detectors measure the electromagnetic 
component and the muonic component separately. 
This enables to reconstruct 
the total number of muons on an event-by-event basis. 
Figure~\ref{fig:kgspec}, left panel shows the 2-dimensional shower 
size (muons and charged particles) 
distribution, which is the basis of the energy and mass investigation at KASCADE-Grande. 

Using the hadronic interaction model QGSJet-II, a composition
independent all-particle energy spectrum was determined in the energy range 
of $10^{16}\,$eV to $10^{18}\,$eV for the Grande data within a total uncertainty in 
flux of 10-15\%~\cite{kgEspec}.
The observables taken into account for the reconstruction are the shower 
size $N_{ch}$ and the muon shower size $N_\mu$. 
Using the reconstruction of the energy spectrum by correlating 
$N_{ch}$ and $N_{\mu}$ on an event-by-event basis, the mass 
sensitivity is minimized by means of a parameter $k(N_{ch},N_{\mu})$, which describes the mass 
normalized ratio of the two observables. 

Despite the overall steeply falling power law behavior of the resulting all-particle spectrum, 
surprisingly, and for the first time, there are some structures observed, which do not allow to 
describe the spectrum with a single slope index. Figure~\ref{fig:kgspec}, right panel shows the 
spectrum in a way visualizing these structures. 
There is a clear evidence that just above $10^{16}\,$eV the spectrum shows a `concave' behavior, 
which is significant with respect to the systematic and statistical uncertainties.   
A further feature in the spectrum is a small break, i.e. knee-like feature at around $10^{17}\,$eV. 
This slope change occurs at an energy where the rigidity dependent, i.e. charge dependent, knee 
of the iron (heavy) component would be expected. 

\begin{figure}[t]
 \includegraphics[width=0.59\textwidth]{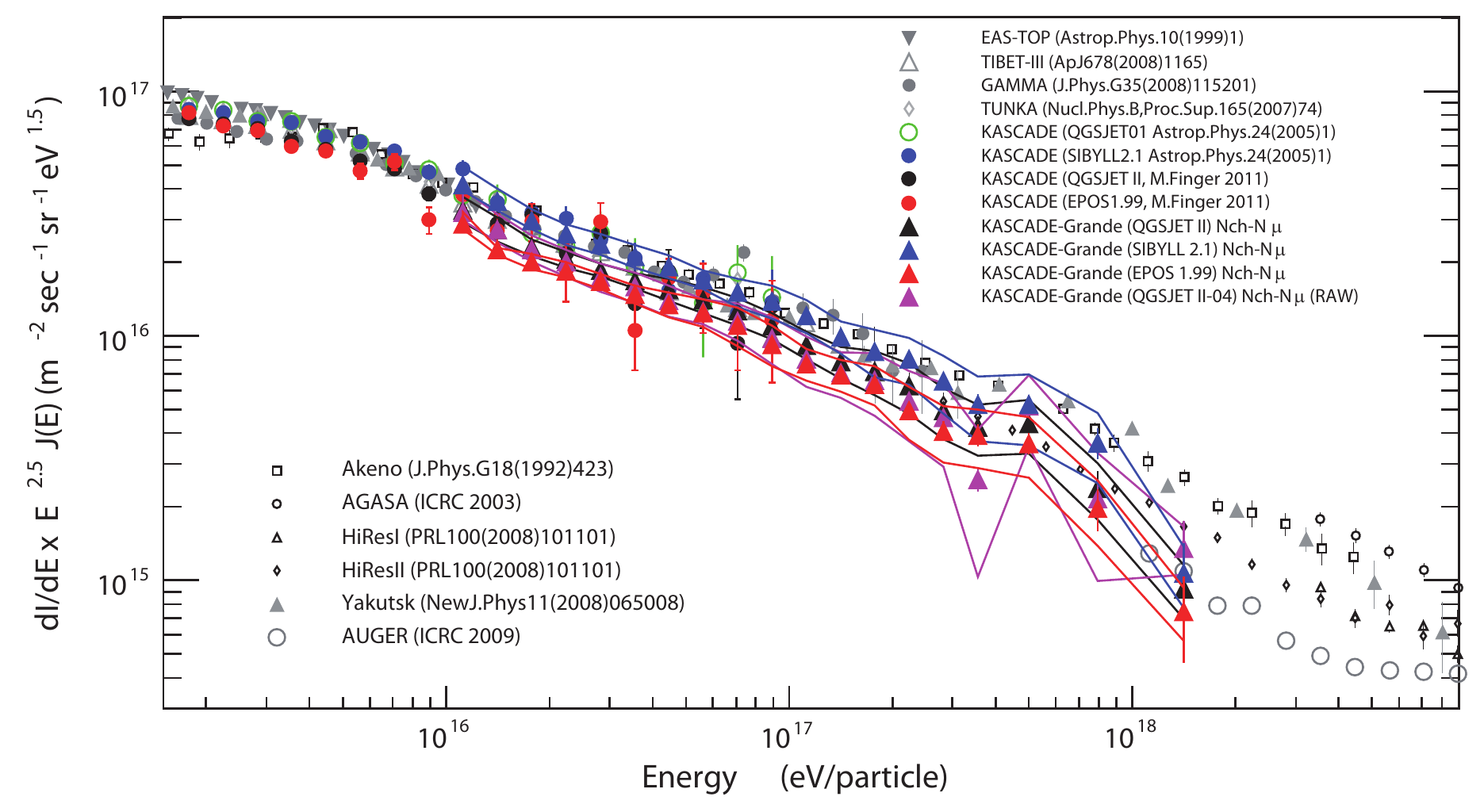} 
 \includegraphics[width=0.39\textwidth]{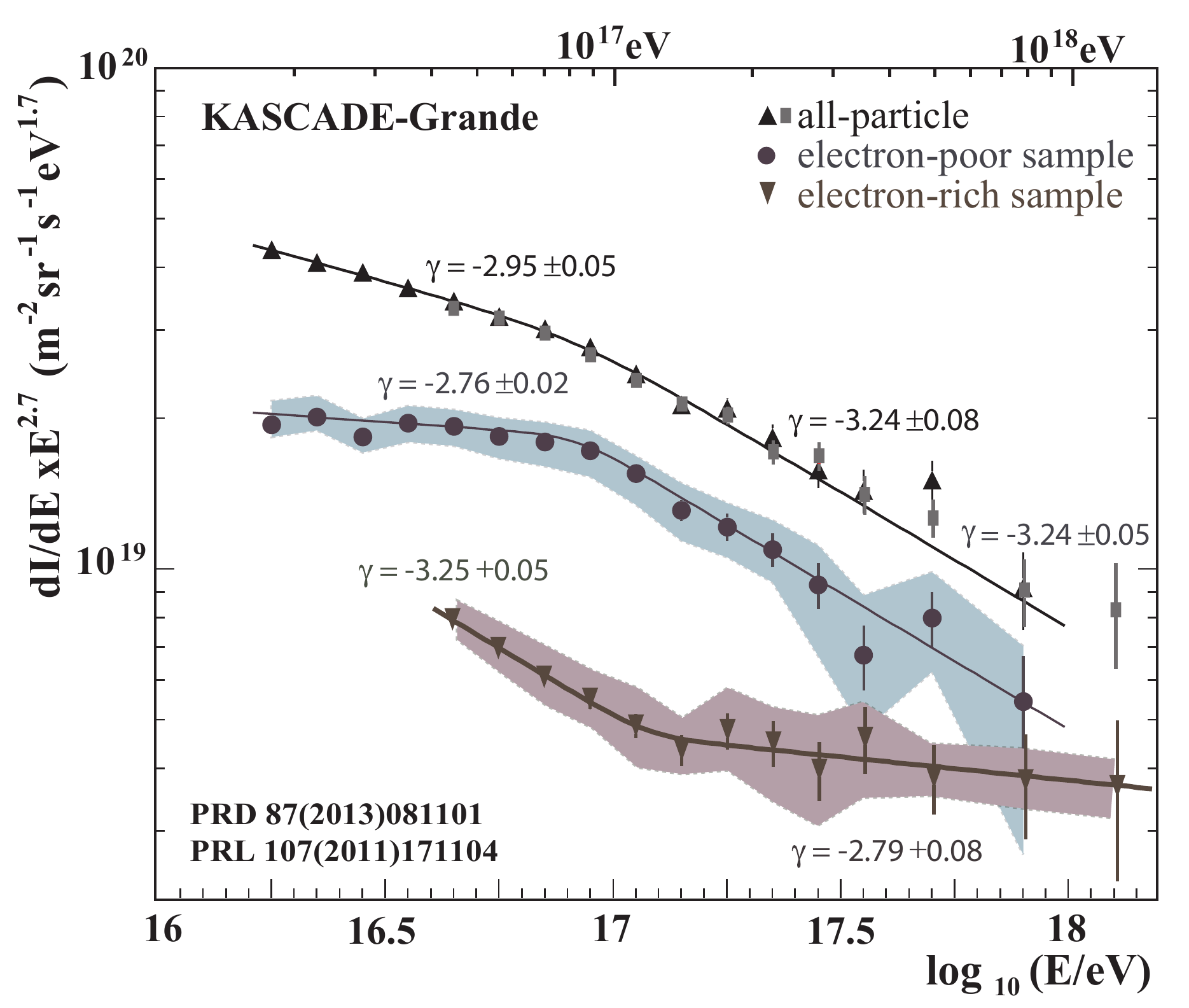} 
\caption{Left panel: The all-particle energy spectrum obtained with KASCADE-Grande 
data based on SIBYLL, QGSJet, QGSJet-II, and EPOS models as well as results of other experiments. 
The band denotes the systematic uncertainties in the flux estimation (from~\cite{icrcmario}).
Right panel: All-particle, light-mass enriched, and heavy-mass enriched energy
spectra from KASCADE-Grande. One all-particle and the
heavy enriched spectra is from one analysis, the other all-particle
and the light primary spectrum result from a larger data set
with higher energy threshold (from~\cite{icrchaungs}).}
\label{fig:kgModels}
\end{figure}

As the calibration of air-shower events always depends on hadronic interaction models 
describing the development of the EAS in the atmosphere, it is crucial for these experiments 
to verify the sensitivity of the observables in use. 
The strategy at KASCADE-Grande is to reconstruct the spectrum on basis of different hadronic 
interaction models. 
It was found that the relative abundances of various mass groups, or in general the analysis of the 
elemental composition, are much more depending on the interaction model in 
use than the all-particle spectrum~\cite{marioJASR}.
Figure~\ref{fig:kgModels}, left panel, shows the obtained spectra reconstructed on 
basis of four different hadronic interaction models.
The structure or characteristics of the spectra are found to be much less affected by the differences of 
the various hadronic interaction models than the relative abundances of the masses. 
Despite the fact, that the discussed spectrum is based on the QGSJet-II hadronic 
interaction model~\cite{qgsjet}, there is confidence that all the found structures of 
the energy spectrum remain stable. 

Not important for the focus of this paper, but for completeness, the recent results 
of KASCADE-Grande on the investigations of the elemental composition will be mentioned in 
the following.
The goal of the KASCADE-Grande experiment is the reconstruction of 
individual mass group spectra. Structures observed in these individual spectra provide 
stronger constraints to astrophysical models of origin and propagation of high-energy cosmic rays than
the all-particle spectrum or a mean logarithmic mass.
For example, already in 2005 KASCADE could prove~\cite{kas-unf} 
that the knee is caused by a strong 
decrease of the light mass group of primary particles and not by heavy primary particles. 
Meanwhile, KASCADE-Grande has investigated such individual mass group spectra also 
at higher primary energies. 
The evolution of the above mentioned $k$ as a function of energy keeps track of the 
evolution of the composition, and allows an event-by-event separation between light, 
medium and heavy primaries, at least. 
Using $k$ as separation parameter for different mass groups, where the normalizations
of $k$ have to be determined with help of simulations, directly the energy spectra of 
the mass groups are obtained~\cite{prl107,Apel2013}. 
All the simulations for the described analyses are performed with the air-shower simulation 
package CORSIKA~\cite{corsika} with QGSJet as hadronic interaction model.  
The application of this methodical approach to shower selection and separation in various 
mass groups is performed and cross-checked in different ways, where the right panel of 
Figure~\ref{fig:kgModels} shows the main results:

The reconstructed spectrum of the electron-poor events, i.e. the spectrum of heavy primaries, shows 
a distinct knee-like feature at about $8 \cdot 10^{16}\,$eV. 
The change of the spectral slope is with $\Delta \gamma = -0.48$ is much larger than at 
the all-particle spectrum with $\Delta \gamma = -0.29$. 
Hence, the selection of heavy primaries enhances the knee-like feature that is already 
present in the all-particle spectrum. 
In addition, an ankle-like feature was found in the spectrum of the electron-rich events, e.g. 
light elements of the primary cosmic rays, at an energy of  $10^{17.08 \pm 0.08} \, \mathrm{eV}$. 
At this energy, the spectral index changes by $\Delta \gamma = 0.46$.

In summary, most important result from KASCADE is the proof that the knee feature at several PeV is 
due to a decrease in the flux of light atomic nuclei of primary cosmic rays. 
Recent results of KASCADE-Grande have now shown two more spectral features: 
a knee-like structure in the spectrum of heavy primaries at around 90 PeV and a hardening of 
the spectrum of light primaries at energies just above 100 PeV.

\subsection{Tunka-133}

\begin{figure}[ht]
\begin{center}
 \includegraphics[width=0.47\textwidth]{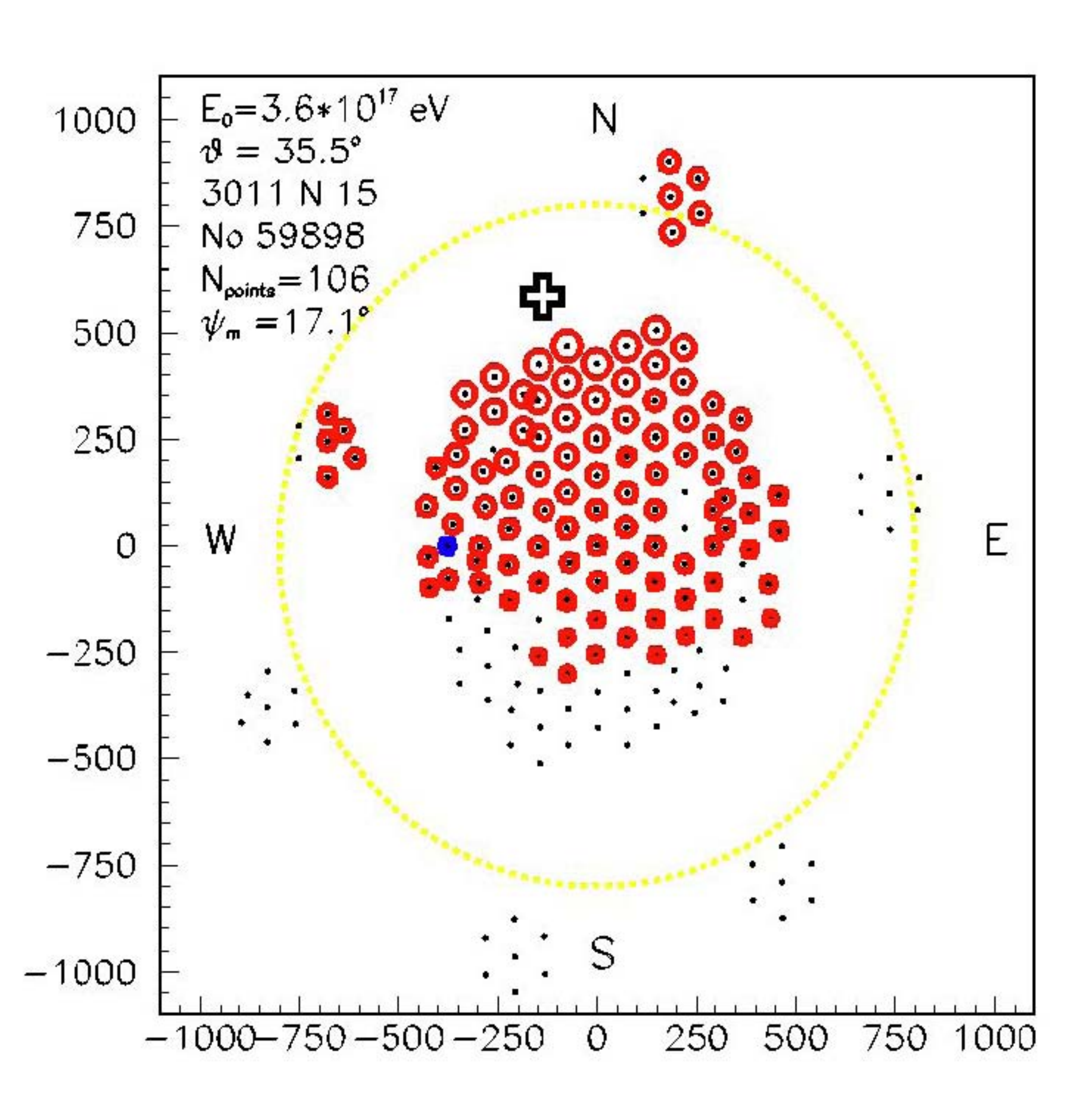} 
 \includegraphics[width=0.50\textwidth]{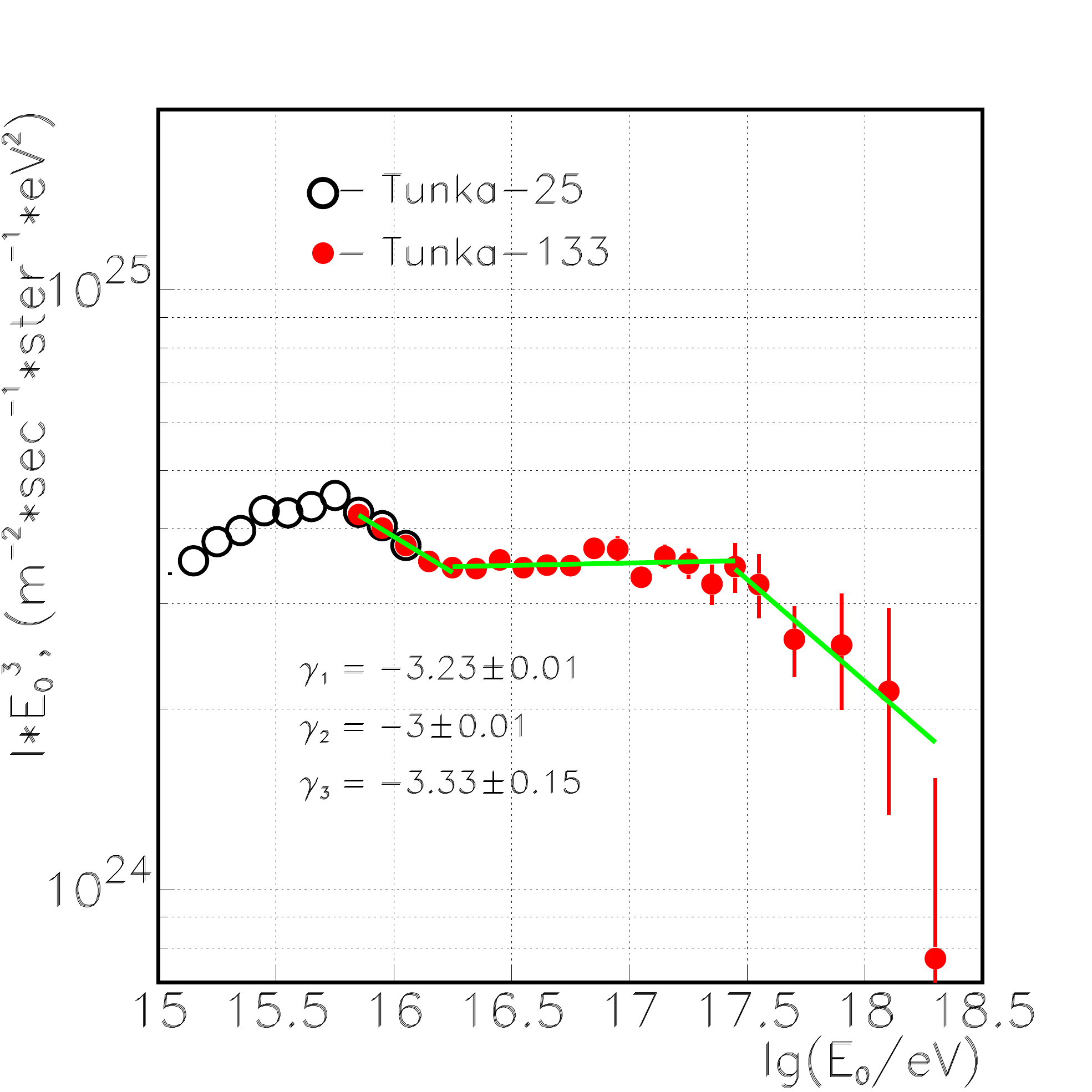} 
\end{center}
\caption{Left panel: The layout of the Tunka-133 installation with one measured example of a typical event.
The size of dots visualizes the amplitude of the detected Cherenkov light at the individual stations. The cross 
shows the estimated shower core. 
Right panel: The primary all-particle energy spectrum by the Tunka-133 
experiment (from~\cite{icrcprosin,nimprosin}).}
\label{fig:tunka}
\end{figure}

To determine the primary energy spectrum and elemental composition the array Tunka-133~\cite{nim12} 
with $\approx 1\,$km$^2$ sensitive area has been installed in the Tunka Valley, Siberia. 
It records the Cherenkov light of EAS using 133 detectors arranged in 19 compact clusters.
The addition of 6 outer clusters in 2011 (see Figure~\ref{fig:tunka}) at 
distances of about 1\,km from the center allows Tunka now to measure with  higher statistics and 
a higher reconstruction quality for events of energies $> 10^{17}$\,eV. 

The signal of the detectors are used to determine the lateral distribution of the Cherenkov light 
emitted by the individual EAS. The characteristic form of this distribution is used to estimate both: 
the primary energy and the primary mass of the incoming air showers.
As a measure of the primary energy the Cherenkov light density at a core
distance of 200\,m, i.e.~$Q(200)$ is used. The connection between the primary energy E$_0$ and 
$Q(200)$ can be expressed by $E_0=C\cdot {Q(200)}^{g}$, where
it was found by CORSIKA simulations, that for the energy range of 
\mbox{$10^{16} - 10^{18}$\,eV} and the zenith angle range of
\mbox{$0^{\circ} - 45^{\circ}$}, as well as a mixed composition (consisting of equal 
contribution of protons and iron nuclei) the value of the index $g$ is 0.94. 
These simulations and the calibration lead to an energy resolution of 15\% for the individual events.

Tunka-133 reconstructs a combined energy spectrum
(Fig.\ref{fig:tunka}, right panel) for events with $R<450$\,m for energies  $< 10^{17}$\,eV
and for events with $R<800$\,m for higher energies. 
The combined spectrum contains about 1900 events with $E_0 > 10^{17}$\,eV.  
This spectrum can not be fitted with a single power law, but with a combination of three power 
laws with different slopes, as indicated in the figure.
The structures in the spectrum are very similar to those by KASCADE-Grande and interestingly, 
by applying a fudge factor of 7\% to the overall flux, the agreement between the two spectra 
is incredibly good.

The analysis of the elemental composition is using the steepness of the lateral distribution 
function, i.e.~the ratio of $Q(100)/Q(200)$, and expressing it in terms of the mean logarithmic mass.
First results are reported in~\cite{nimprosin} and are roughly in agreement with the results obtained
at KASCADE-Grande.

\subsection{IceTop}

\begin{figure}[ht]
\begin{center}
 \includegraphics[width=0.35\textwidth, height=0.32\textwidth]{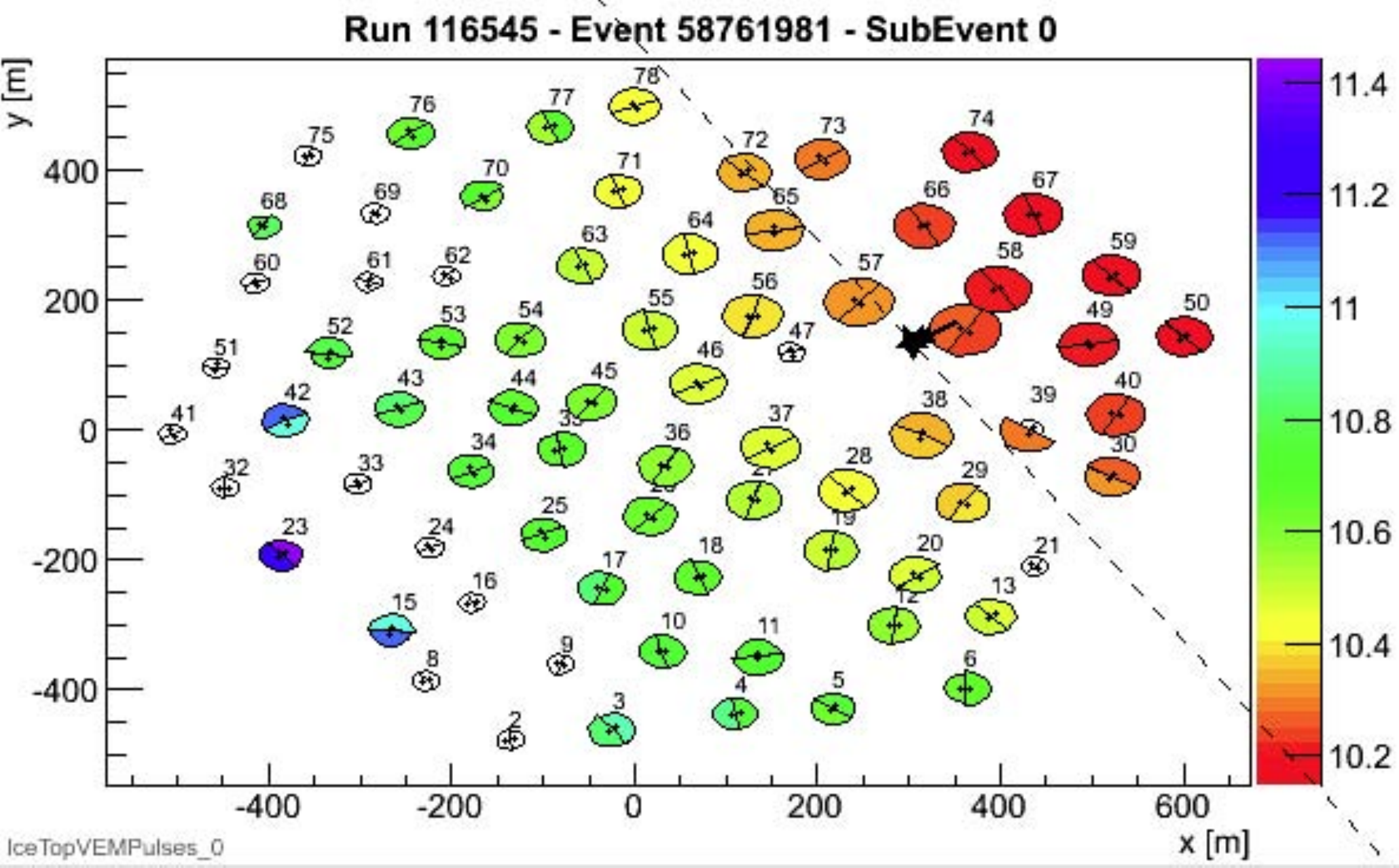}
 \hspace*{0.02\textwidth} 
 \includegraphics[width=0.59\textwidth]{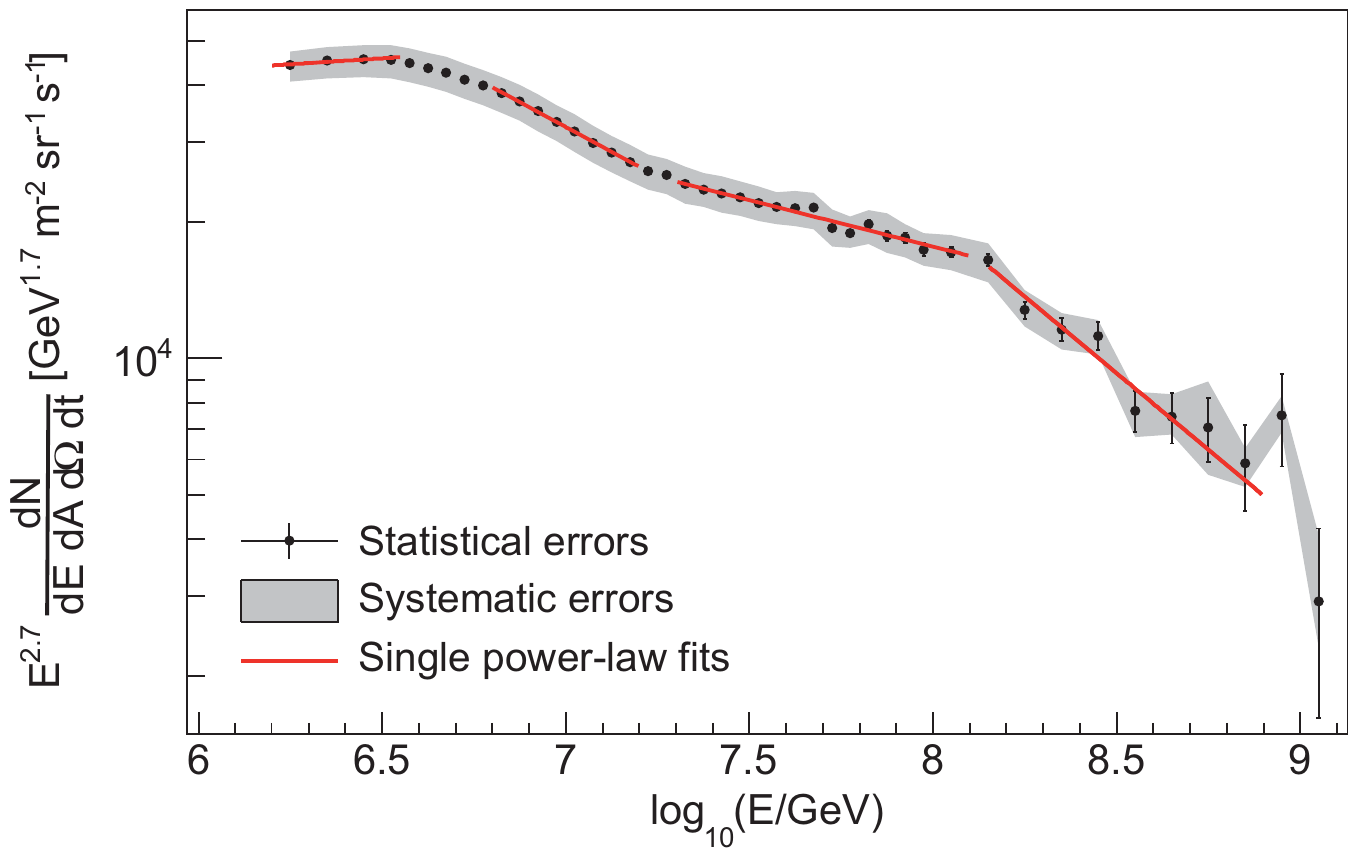} 
\end{center}
\caption{Left panel: The layout of the IceTop installation with a typical measurement.
The size of dots visualizes the amplitude of the detected particle density at the individual stations; 
the color code marks the measured arrival times. 
Right panel: The all-particle cosmic ray spectrum obtained by IceTop, where the shaded area represents the systematic errors. Spectral fits in different energy ranges are indicated (from~\cite{icetopspec}).}
\label{fig:icetop}
\end{figure}

IceTop~\cite{IT_detP} is the surface array of the IceCube Neutrino Observatory. 
IceTop detects air showers from primary cosmic rays in the 300\,TeV to 1\,EeV energy range. 
The array will consist of 81 surface stations in its final configuration covering an area of 
one square kilometer (see Figure~\ref{fig:icetop}). 
Each station consists of two ice-Cherenkov tanks separated by 10\,m. Due to the relative 
high-altitude observation level, the IceTop detector stations record mainly the signal of the 
electromagnetic component of the air shower. 
The recorded signal in the stations are calibrated in terms of 'Vertical Equivalent Muons' (VEM).
The IceTop reconstruction algorithm uses information from individual tanks, including location, measured light and pulse time. Shower direction, core location and shower size are reconstructed by fitting 
the measured charges with a Lateral Distribution Function (LDF) and the signal times with a function 
describing the geometric shape of the shower front. 
The relevant parameter for the energy determination, the shower size, S$_{125}$, is defined as the 
fitted value of the LDF at a perpendicular distance of 125\,m away from the shower axis. 

Recently, the reconstructed spectrum was reported in the region from 1.6\,PeV up to 1.3\,EeV based 
on S$_{125}$ by the IceTop air shower array in its 73 station configuration. 
When showing spectra for a given zenith range and assumed composition, the energy is 
estimated with an appropriate calibration function, where the parameters are obtained with help of simulations. 
In addition, the S$_{125}$  to energy conversion is performed assuming a mixed primary composition 
as described in reference~\cite{H4a} and is referred to as H4a model. 

The spectrum to be discussed here (see Figure~\ref{fig:icetop}, right panel) was derived assuming 
the H4a model and averaging over the full zenith 
range up to $37^\circ$. The analysis is based on the hadronic interaction model SIBYLL~\cite{sibyll}.
The shown spectrum includes an unfolding in which the spectrum derived in the previous step is used 
to determine the effective area and the $S_{125}$-to-$E_{true}$ relation for the next spectrum evaluation. 
In case of convergence the effective area effectively accounts for migrations due to finite resolutions.
Also IceTop observes that the all-particle cosmic-ray energy spectrum does not follow a single power 
law above the knee, but shows significant structure.  
Hence, the spectrum was fitted by simple power functions in four different energy ranges 
(see also Figure~\ref{fig:icetop}) with three structures:
The knee, where the index changes smoothly between 4 to 7\,PeV; a hardening at around $18\pm2$\,PeV; and a 
sharp fall is observed beyond $130\pm30$\,PeV. 
The hardening as well as the steepening are clear signatures of the spectrum and can not be attributed 
to any of the systematics or detector artefacts~\cite{icetopspec}.

In case of the IceTop spectrum reconstruction the difference in the spectra obtained using
SIBYLL or QGSJet as an interaction model are
much smaller than in case of KASCADE-Grande. This probably is due to the
much lower altitude of the KASCADE-Grande detector.

In a later stage, IceTop measurements will be combined with the signal of high-energy muons measured with 
the in-ice IceCube installation and/or low-energy muons measured by IceTop stations at large 
distances to the shower core in order to estimate the elemental composition.

\section{Discussion of the All-Particle Spectra}

\begin{figure}[ht]
\begin{center}
 \includegraphics[width=0.75\textwidth]{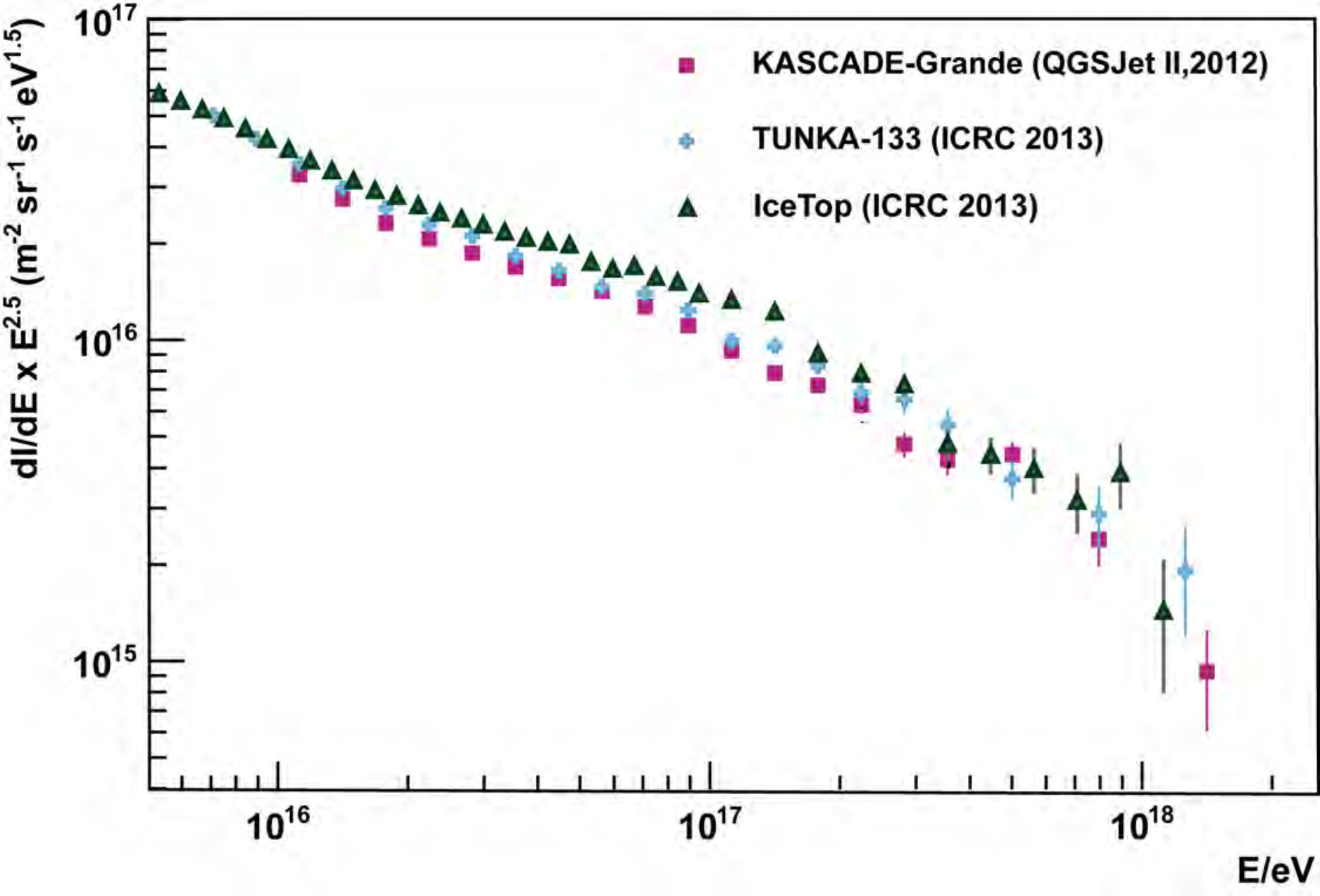} 
\end{center}
\caption{The all-particle, cosmic-ray energy spectrum by three different experiments(KASCADE-Grande~\cite{kgEspec};
Tunka-133~\cite{nimprosin}; IceTop~\cite{icetopspec}).}
\label{fig:spectra}
\end{figure}

Despite the overall smooth power-law behavior of the all-particle spectrum 
(see Fig.~\ref{fig:spectra}), there are some structures observed, which do not allow 
to describe the spectrum with a single slope index. 
Of course, these structures are smaller than the well-known knee or ankle features, 
but statistically significant, and identified by all three experiments KASCADE-Grande, Tunka-133, IceTop. 

There is a clear evidence that just above $10^{16}\,$eV the spectrum shows a `concave' behavior; 
i.e.~a hardening of the spectrum appears.    
This is observed by all three experiments with high statistical accuracy, despite the fact that KASCADE-Grande 
report the spectrum above $10^{16}\,$eV, only, and Tunka-133 has a change of experimental configuration 
from Tunka-25 to Tunka-133 roughly in this energy range.   

Another feature, also seen by all three experiments, is a small break at around $10^{17}\,$eV, which 
appears at KASCADE-Grande at a little lower energy than at the other two experiments. 
Applying a second power law a statistical significance increase of the index is similarly obtained at all 
three experiments. That means, that this second knee close to 100 PeV, first identified and 
reported by KASCADE-Grande, is now established due to the experimental confirmation by Tunka-133 and IceTop.

KASCADE-Grande has shown that the spectral form does not depend on the hadronic interaction model in use.
This is not true for the absolute flux, where a difference of 15-20\% can appear. As KASCADE-Grande measures the 
total number of electrons and muons, separately, these differences are related to the absolute normalization 
of the energy scale by the various models.
IceTop has reported results on the basis of SIBYLL and QGSJet, but found the difference very small, which is 
probably owned to the observation level close to the shower maximum. 
Tunka-133's measurements are somehow based on a calorimetric method; i.e.~the energy calibration is seen as 
less dependent from the hadronic interaction model. This statement still needs to be confirmed by 
further investigations of the Tunka data.  

The main difference of the three spectra is the absolute normalization to the energy scale, 
which moves the found structures slightly in energy and by that also in the absolute flux. 
But the spectra agree nicely within the given systematic uncertainties.  
The normalizations depend on the calibration (hadronic interaction model) as well as on the 
treatment of including the a-priori unknown elemental composition.  
Hence, a source of the differences in the spectra is due to assumptions in the 
composition. All experiments take this into account in the estimation of the systematic uncertainties, 
but when we discuss the differences in the flux of the 
all-particle spectra obtained by the three experiments this contribution can not be resolved easily.

Despite the fact that the three experiments use different observation techniques, are located at 
different observation levels, and use different hadronic interaction models to interpret their data, 
\begin{figure}[ht]
\begin{center}
 \includegraphics[width=0.95\textwidth]{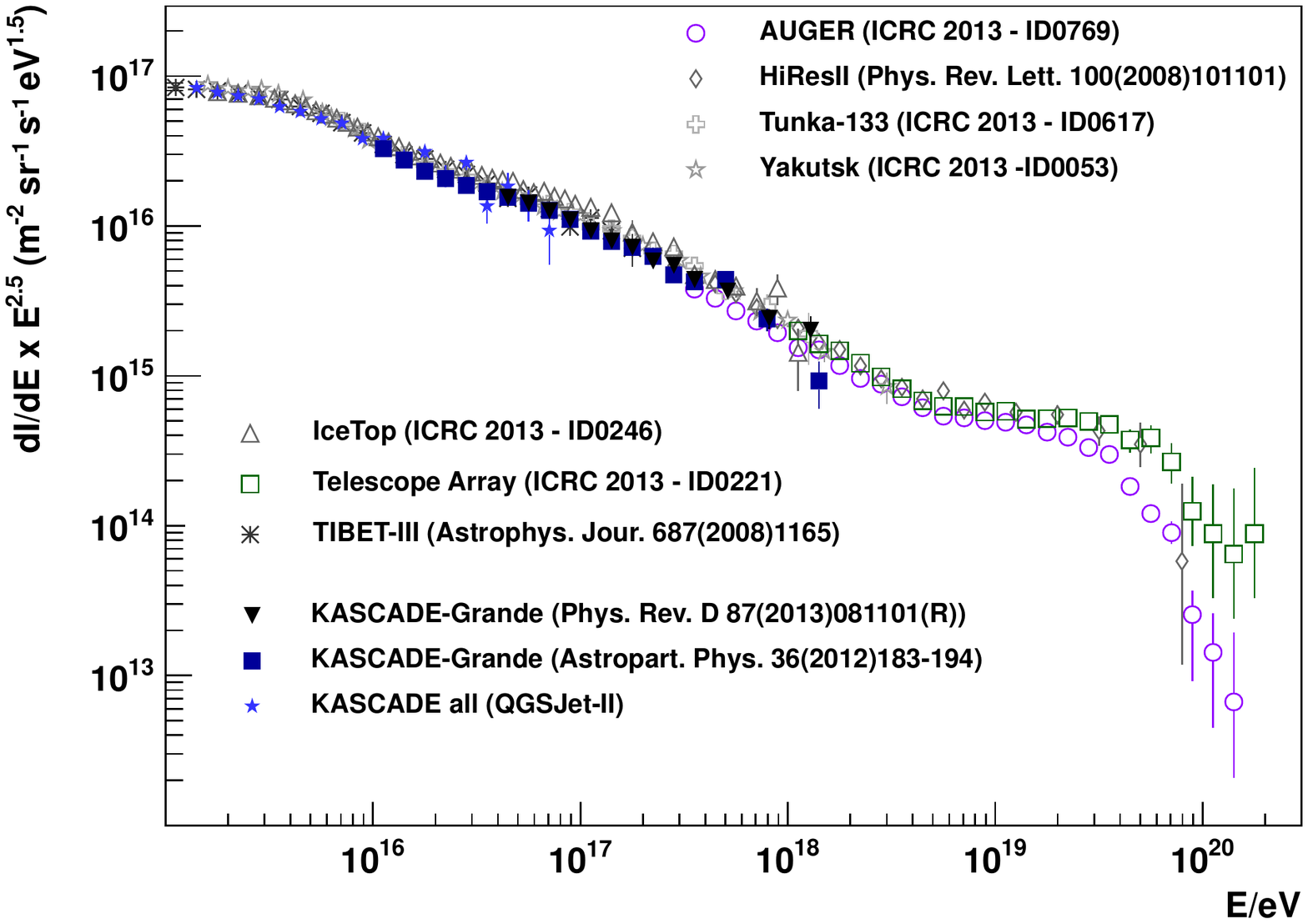} 
\end{center}
\caption{Comparison of the all-particle energy spectra obtained with KASCADE-Grande, Tunka-133,
and IceTop-71 to a wider energy range and results of other experiments.}
\label{fig:allspec}
\end{figure}
the agreement within 15\% of the total flux is surprisingly good.
This on one hand confirms the structures found so far, but on the other hand also confirms the 
high quality of the data taken by these modern experiments and also the validity of the hadronic 
interaction models (at least for this energy range and for the observables measured).
But, as the experiments identified the same structures and the absolute energy normalization depends on the 
hadronic interaction models, we can cross-check these models by applying the same composition 
assumptions and the same hadronic model to the individual reconstruction procedures. 
Such a study should lead to a very detailed cross-check of the models as the three experiments have different
observables at different observation levels.
Discussions in this direction between the three collaborations have already started and will be continued.  
   
Figure~\ref{fig:allspec} compiles the energy spectra discussed with results of other experiments.
Despite the independent measurements and data analysis there is at low energies a very good agreement 
with the results of the KASCADE experiment  and others in the overlapping energy range. 
At higher energies the KASCADE-Grande spectrum (QGSJet~II) shows a slightly lower flux than earlier 
experiments and the IceTop and Tunka-133 results. 
At the highest accessible energy the KASCADE-Grande, IceTop, and Tunka-133 results are statistically 
in agreement with the results of HiRes and the Pierre Auger Observatory~\cite{icrcPAO}.

\section{Implication for Astrophysical Models}

The position of the well established knee is roughly in agreement with the energy where 
supernova remnants (SNR) become inefficient accelerating particles~\cite{hillas} under the 
assumption that this acceleration increases proportionally to the charge of the cosmic particles.
Using this standard picture, various theories with different assumptions were then developed 
to explain the behavior of the spectrum between the knee and ankle features. 
The basic idea of the `dip model' is that the ankle is a 
feature of the propagation of extragalactic protons. 
Consequently, in that model the composition at the ankle is to a large extent proton-dominant 
and the transition from galactic to extragalactic origin of cosmic rays occurs already at 
energies well below $10^{18}\,$eV.   
In the scenario of the dip model, at energies around 
$10^{17}\,$eV a pure galactic iron component should be left. Consequently the transition occurs 
already at energies  well below the ankle. 
On the other hand, to avoid an early appearance of the extragalactic cosmic ray component (which 
can be of pure proton~\cite{berezinsky}, but also of mixed composition~\cite{allard}),
Hillas~\cite{hillas} proposed in addition to the standard SNR component, a `component B' 
of cosmic rays of (probably) galactic origin. 
This component would also experience a charge dependence 
of break-offs, but now shifted to approximately ten times higher in energy. 
As a result, the transition occurs here at the ankle and for the entire energy range from 
$10^{15}\,$eV to $10^{18}\,$eV a mixed elemental composition is expected. 
In this scenario, the second knee would be a feature of the component B. 
This is also the basic idea of the H4-model~\cite{H4a} used as composition assumption in the 
IceTop reconstruction procedures.  

What follows is a short confrontation of the experimental results discussed above with these
astrophysical models:
The hardening of the spectrum is expected when a pure rigidity dependence of the 
galactic cosmic rays is assumed. Depending on the relative abundances of the different primaries 
one would expect charge dependent steps in the all-particle spectrum.    
The gap in the knee positions of light primaries (proton, helium, and CNO group 
of $Z=1-8$) and the heavy group can lead to a hardening of the spectrum~\cite{donato}. 
On the other hand a transition from one source population to another one should 
also result in a hardening of the spectrum. 
In this aspect, the  hardening could be a first 
experimental hint to the `component B' of galactic cosmic rays as proposed by Hillas~\cite{hillas}.

The slope change of the new `second knee' occurs at an energy where the rigidity dependent knee 
of the iron component is expected. 
But, the change of the spectral index is small compared to what has been seen in case 
of protons and helium (the knee), which could be explained 
when the iron component, and in particular the iron component from the standard SNR population, 
is not dominant around $10^{17}\,$eV. This again can happen
in presence of a `component B' of mixed composition.
Despite it is not visible at the all-particle spectrum, for completeness it should be mentioned 
that the found `light ankle' in the spectrum of light primaries by KASCADE-Grande~\cite{Apel2013} 
at energies well below $10^{18}\,$eV also fits in the models favoring a galactic component B. 

A significant conclusion, however, is not possible without investigating the composition in 
detail in this energy range.

\section{Conclusions}

The all-particle cosmic-ray energy spectrum in the energy range from the knee to the ankle 
could recently be reconstructed by three modern experiments (KASCADE-Grande, Tunka-133, IceTop).
Despite measuring different air-shower observables, detecting air-showers at different 
observation levels, and using different hadronic interaction models underlying 
the analysis the spectra are in a good agreement. In particular, all of the experiments 
have identified significant structures in the spectrum.
Namely, a hardening of the all-particle energy spectrum is observed at $\approx 10-20\,$PeV, and a 
small break-off at $\approx 100\,$PeV. 
These features give interesting hints to the astrophysical processes in the transition 
region from galactic to extragalactic origin of cosmic rays. 
The low energy extension of the Pierre Auger Observatory will also contribute in near future 
high-quality measurements to the energy range below the ankle.  

A wealth of information on individual showers is available with the KASCADE-Grande, Tunka-133 and IceTop 
measurements. This will make it possible to reconstruct the all-particle energy spectrum with high precision, 
as well as to investigate in near future the elemental composition, to test the hadronic interaction models, 
and to study cosmic ray anisotropies. KASCADE-Grande has already published first results in this direction.
In addition, discussions between the collaborations have started to combine the information of all three 
installations for even more stringent tests on composition and hadronic interaction models.  

\section*{Acknowledgements}
The author acknowledges the colleagues from the KASCADE-Grande, Tunka-133, and IceTop Collaborations 
for the fruitful discussions on the topic. In particular, many thanks to Tom Gaisser, Javier Gonzalez, 
and Bakhtiyar Ruzybayev from IceTop; Vasily Prosin and Leonid Kuzmichev from Tunka-133; as well as 
Sven Schoo, Donghwa Kang, and Mario Bertaina from KASCADE-Grande.  
This work was partially supported by the `Helmholtz Alliance for Astroparticle Physics
HAP' funded by the Initiative and Networking Fund of the Helmholtz
Association, Germany.

\bibliographystyle{elsarticle-num}

\end{document}